\documentclass[referee,useAMS,usenatbib,usegraphicx]{mn2e}

\newcommand{\saxj}{SAX~J1808.4$-$3658}
\newcommand{\igrj}{IGR~J00291+5934}

\def\msun{{\rm M}_\odot}
\def\Porb{P_{\rm orb}}

\def\T0{T^*_0}

\usepackage{subfigure}
\bibpunct{(}{)}{;}{a}{}{,}

\title[Orbital Evolution of an Accreting Millisecond Pulsar]{Orbital 
Evolution of an Accreting Millisecond Pulsar: Witnessing the Banquet of a 
Hidden Black Widow?}

\author[T. Di Salvo et al.]{T. Di Salvo$^{1*}$, 
L. Burderi$^2$\thanks{E-mail:disalvo@fisica.unipa.it, burderi@mporzio.astro.it}, 
A. Riggio$^2$, A. Papitto$^{3,4}$, M.T. Menna$^3$\\
$^1$Dipartimento di Scienze Fisiche ed Astronomiche,
Universit\`a degli Studi di Palermo, via Archirafi 36 - 90123 Palermo, Italy\\
$^2$Dipartimento di Fisica, Universit\`a degli Studi di Cagliari, 
SP Monserrato-Sestu, KM 0.7, 09042 Monserrato, Italy \\
$^3$I.N.A.F. - Osservatorio Astronomico di Roma, via Frascati 33,
00040 Monteporzio Catone (Roma), Italy\\
$^4$Dipartimento di Fisica, Universit\`a degli Studi di Roma 
`Tor Vergata', via della Ricerca Scientifica 1, 00133 Roma, Italy}

\begin{document}

\date{}

\maketitle

\begin{abstract}
We have performed a timing analysis of all the four X-ray outbursts from the
accreting millisecond pulsar \saxj\ observed so far by the PCA on board 
RXTE. For each of the outbursts we derived the best-fit value of the 
time of ascending node passage. We find that these times follow a 
parabolic trend, which gives an orbital period 
derivative $\dot P_{\rm orb} = (3.40 \pm 0.18) \times 10^{-12}$ s/s, and 
a refined estimate of the orbital period, $P_{\rm orb} = 7249.156499 
\pm 1.8 \times 10^{-5}$ s (reference epoch $T_0 = 50914.8099$ MJD).
This derivative is positive, suggesting a degenerate or fully convective 
companion star, but is more than one order of magnitude higher than what is 
expected from secular evolution driven by angular momentum losses caused by 
gravitational radiation under the hypothesis of conservative mass transfer. 
Using simple considerations on the angular momentum of the system,
we propose an explanation of 
this puzzling result assuming that during X-ray quiescence the source is 
ejecting matter (and angular momentum) from the inner Lagrangian point. 
We have also verified that this behavior is in agreement with a possible 
secular evolution of the system under the hypothesis of highly non-conservative
mass transfer. In this case, we find stringent constraints on the masses of the 
two components of the binary system and its inclination.
The proposed orbital evolution indicates that in this kind of 
sources the neutron star is capable to efficiently ablate the companion star, 
suggesting that this kind of objects are part of the population of the so-called 
black widow pulsars, still visible in X-rays during transient mass accretion episodes.
\end{abstract}


\section{Introduction}

\saxj\ is the first discovered among the ten known accreting 
millisecond pulsars (hereafter AMSPs), which are all transient X-ray 
sources, and is still the richest laboratory for timing studies of 
this class of objects. Although timing analysis have been now 
performed on most of the sources of this sample with interesting 
results (see Di Salvo et al.\ 2007 for a review and references therein),
\saxj\ is the only known AMSP for which several outbursts have been 
observed by the RXTE/PCA with high time resolution. 
In particular, the first outburst of this source was  observed by the RXTE/PCA in April 1998, when coherent X-ray pulsations at $\sim 2.5$ ms 
and orbital period of $\sim 2$~hr (Wijnands \& van der Klis 1998;
Chakrabarty \& Morgan 1998) were discovered. The source showed other
X-ray outbursts in 2000 (when only the final part of the outburst could 
be observed, Wijnands et al.\ 2001), in 2002 (when kHz QPOs and 
quasi-coherent oscillations during type-I X-ray bursts were discovered, 
Wijnands et al.\ 2003; Chakrabarty et al.\ 2003), and again in 2005, 
approximately every two years (see Wijnands 2005 for a review).

Although widely observed, \saxj\ remains one of the most enigmatic
sources among AMSPs, since timing analyses performed on this source
have given puzzling results. Burderi et al.\ (2006), analysing the
2002 outburst, found that the pulse phases (namely the pulse arrival 
times) show evident shifts, probably caused by variations of the pulse 
profile shape. They noted that the phases derived from the second 
harmonic of the pulse profile were much more stable, and tentatively 
derived a spin frequency derivative from these data. 
To explain the relatively large frequency derivative a quite high mass 
accretion rate was required, about a factor of 2 higher than the 
extrapolated bolometric luminosity of the source during the same 
outburst. Hartman et al.\ (2008, hereafter H08) performed a
timing analysis of all the four outbursts of \saxj\ observed up to
date by RXTE, finding again complex phase shifts in all of them. Their conclusion was that the large variations of the pulse
shape do not allow to infer any spin frequency evolution during a single
outburst, with typical upper limits of $|\dot \nu| \la 2.5 \times
10^{-14}$ Hz/s ($95\%$ c.l.), which were derived excluding the first 
few days of the 2002 and 2005 outbursts and the large residuals at the 
2002 mid-outburst. Interestingly, combining the results from all
the analysed outbursts, H08 found a secular spin
frequency derivative of $\dot \nu = (-5.6 \pm 2.0) \times 10^{-16}$
Hz/s, indicating a secular spin-down of the neutron star in this system.
From this measure they found an upper limit of $1.5 \times 10^8$ Gauss 
to the neutron star magnetic field.

Papitto et al.\ (2005) performed a temporal analysis of the outbursts
of \saxj\ that occurred in 1998, 2000, and 2002, which resulted in 
improved orbital parameters of the system. The large
uncertainty caused by the relatively limited temporal baseline made it
impossible to derive an estimate of the orbital period derivative.
In this paper we use all the four outbursts of \saxj\ observed by RXTE/PCA, 
spanning a temporal baseline of more than 7 years, to derive an orbital 
period derivative, the first reported to date for an AMSP. 
The value we find with high statistical significance
is surprising, $\dot P_{\rm orb} = (3.40 \pm 0.18) \times 10^{-12}$ s/s. 
This value for the
orbital period derivative is compatible with the measure reported by 
H08, which, independently and simultaneously, have found 
the orbital period derivative of \saxj. In \S~3 we propose a simple 
explanation of this result arising from considerations on the conservation 
of the angular momentum of the system, 
which is consistent with a (non-conservative) secular evolution of 
the system. In particular, we hypothesize that during quiescence \saxj\ 
experiences a highly non-conservative mass transfer, in which a great 
quantity of mass is lost from the system with a relatively high specific 
angular momentum.

\section{Timing Analysis and Results}

In this paper we analyse RXTE public archive data of \saxj\ taken during 
the April 1998 (Obs.\ ID P30411), the February 2000 (Obs.\ ID P40035), the 
October 2002 (Obs.\ ID P70080), and the June 2005 (Obs.\ ID P91056 and 
Obs.\ ID P91418) outbursts, respectively (see Wijnands 2005 and H08 for 
a detailed description of these observations). 
In particular, we analysed data from the PCA (Jahoda et al. 1996),
which is composed of a set of five xenon proportional counters
operating in the $2-60$ keV energy range with a total effective area of
6000 cm$^2$. For the timing analysis, we used event mode data with 64
energy channels and a 122 $\mu$s temporal resolution. We considered only
the events in the $3-13$ keV energy range where the signal to noise ratio 
is the highest, but we checked that a different choice (considering  
for instance the whole RXTE/PCA energy range) does not change the results
described below. The arrival times of
all the events were referred to the solar system barycenter by using 
JPL DE-405 ephemerides along with spacecraft ephemerides. This task 
was performed with the faxbary tool, considering as the best estimate for 
the source coordinates the radio counterpart position, that has a $90\%$ 
confidence error circle of 0.4 arcsec radius, which is compatible with 
that of the optical counterpart (Rupen et al.\ 2002; Giles et al.\ 1999).

For each of the outbursts we derived a precise orbital solution using 
standard techniques (see e.g.\ Burderi et al.\ 2007; Papitto et al.\ 2007,
and references therein). In particular, we firstly corrected the arrival
times of all the events with the orbital solution given by Papitto et al.\
(2005). Then we looked for differential corrections to the adopted orbital
parameters as described in the following. We epoch-folded time intervals
with a duration of about 720~s each ($1/10$ of the orbital period) at the 
spin period of 2.49391975936 ms, and fitted each pulse profile obtained 
in this way with sinusoids in order to derive the pulse arrival times or 
pulse phases. The folding period was kept constant for all the outbursts.
In fact, the determination of the pulse phases is insensitive to the exact 
value of the period chosen to fold the light curves providing that 
this value is not very far from the true spin period of the pulsar. 
Note that H08 have measured a secular derivative of the spin frequency in 
\saxj, that is $\dot \nu = (-5.6 \pm 2.0) \times 10^{-16}$ Hz/s. This is 
a quite small value, and they do not find evidence of variations of the 
spin frequency during a single outburst. 
In order to choose a value of the spin period as close 
as possible to the true one, we used the value above, that is in between the 
best-fit spin period reported by Chakrabarty \& Morgan (1998) for the 1998 
outburst and the best-fit spin period reported by Burderi et al.\ (2006) for 
the 2002 outburst. As in Burderi et al.\ (2006), we fitted each
pulse profile with two sinusoids of fixed periods (the first, with period 
fixed to the spin period adopted for the folding, corresponding to the 
fundamental, and the second, with period fixed to half the spin period, 
corresponding to the first overtone, respectively). 
In all cases the $\chi^2 / d.o.f.$ obtained from the fits of the pulse 
profiles was very close to (most of the times less than) 1. 
In order to improve the orbital solution we used the phase delays of the 
fundamental of the pulse profile which has the best statistics; the 
uncertainties on these phases were derived calculating 1-$\sigma$ statistical 
errors from the fit with two sinusoids. 

We then looked for differential corrections to the adopted orbital parameters,
which can be done by fitting the pulse phases as a function of time for each 
outburst. In general, any residual orbital modulation is superposed to a 
long-term variation of the phases, e.g.\ caused by a variation of the
spin. However, as noted by Burderi et al.\ (2006) for the 2002 outburst, 
\saxj\ shows a very complex behavior of the pulse phases with time, with 
phase shifts, probably caused by variations of the pulse shape, that are
difficult to model and to interpret. To avoid any fitting of this complex 
long-term variation of the phases, we preferred to restrict the fit of 
the differential corrections to the orbital parameters to intervals in 
which the long-term variation and/or shifts of phases are negligible.
We therefore considered consecutive intervals with a duration of at least 
4 orbital periods (depending on the statistics), and fitted the phases of 
each of these intervals with the formula for the differential corrections 
to the orbital parameters (see e.g.\ Deeter, Pravdo \& Boynton 1981 and 
eq.~(3) in Papitto et al.\ 2007). The selection of the intervals is
somewhat arbitrary; we have verified, however, that the results do not
change using a different choice.
No significant corrections were found on the adopted values of the orbital 
period, $P_{orb}$, the projected semimajor axis of the neutron star 
(NS) orbit, $a_1 \sin i / c$, and the eccentricity of the orbit. In particular,
for the eccentricity we find an upper limit of $4.6 \times 10^{-5}$ 
(95\% c.l.) combining all the data. 

\begin{figure}
  \begin{center}
    \begin{minipage}{80mm}
      \includegraphics[width=80mm]{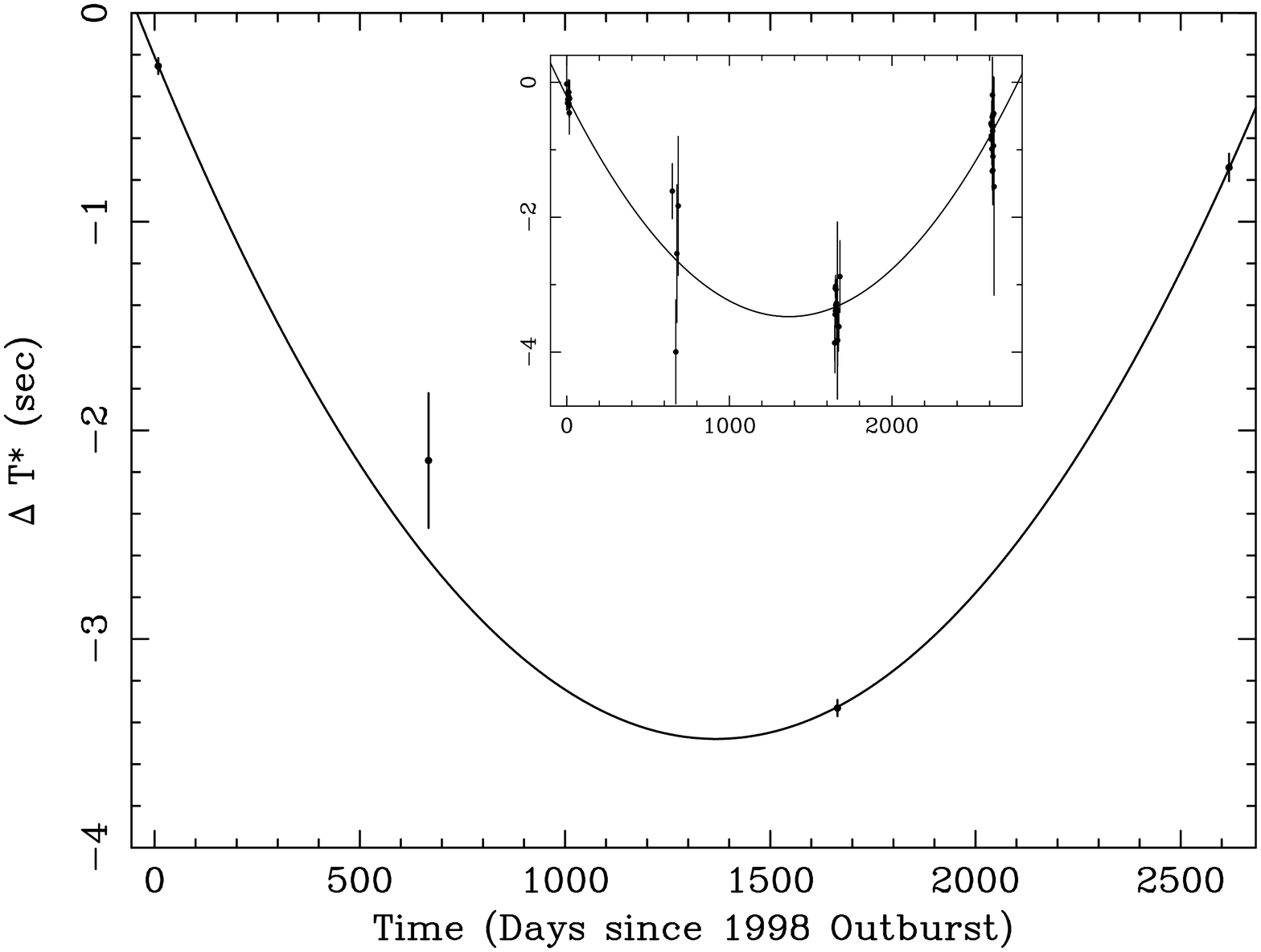}
      \caption{Differential correction, $\Delta T^*$, to the time of passage of 
the NS at the ascending node for each of the four outbursts analysed. 
In the inset we show the single measurements of $\Delta T^*$ obtained for each of
the consecutive time intervals in which each outburst has been divided (see text).
All the times are computed with respect to the beginning of the 1998 outburst,
$T_0 = 50914.8099$ MJD.}
      \label{fig1}
    \end{minipage}
  \end{center}
\end{figure}

On the other hand we found that the times of
passage of the NS at the ascending node at the beginning of each outburst,
$T^*_N$, were significantly different from their predicted values, 
$T^*_0 + N \Porb$ where $T^*_0$ is the adopted time of ascending node passage 
at the beginning of the 1998 outburst and the integer $N$ is the exact 
number of orbital cycles elapsed between two different ascending node 
passages, i.e.\ $N$ is the integer part of $(T_N^{*} - T^*_0)/ P_{orb}$ under 
the assumption that $| T_N^{*} - (T^*_0 + N \Porb) | < P_{orb}$ that we have 
also verified {\it a posteriori}. 
We therefore fixed the values of $P_{orb}$, $a_1 \sin i / c$, and the 
eccentricity, and derived the differential corrections, $\Delta T^*$, 
to the time of passage at the ascending node, obtaining a cluster of points 
for each outburst, which are plotted as a function of time in the inset of 
Figure~\ref{fig1}.  This has been done in order to check that (systematic) 
uncertainties on the arrival  times of  the pulses (such as phase shifts or other 
kind of noise) not already included in our estimated uncertainties for the pulse phases, 
did not significantly affect the determination of the orbital parameters. 
Since each of the points corresponding to an outburst may 
be considered like an independent estimate of the same quantity, if the errors 
on the phases were underestimated, we would expect that errors on the derived 
orbital parameters were underestimated. In this case the scattering of the 
points representing the time of passage at the ascending node derived for each 
of the considered intervals would be larger than the errors associated with each point.
This indeed is not the case; in the inset of Fig.~\ref{fig1}, we have shown 
the results obtained from each of these intervals, and the scattering of the 
points appears always comparable to the associated 1-$\sigma$ error. The largest
scattering is observed for the outburst of 2000, which indeed is the point 
with the largest distance from the best fit parabola (see below). We think 
that this is caused by the worse statistics during this observation, which was 
taken at the end of the 2000 outburst.

We hence combined all the measurements corresponding to each outburst computing  
the error-weighted mean of the corresponding points, obtaining the four points 
shown in Figure~\ref{fig1}.
These points show a clear parabolic trend that we fitted to the formula:
\begin{equation}
\Delta T^* = \delta T^*_0 + \delta P_{orb} \times N + (1/2) 
\dot P_{orb} P_{orb} \times N^2
\label{eq1}
\end{equation}
In this way we found the best fit values $T^*_0 + \delta T^*_0$, 
$\Porb + \delta \Porb$ and $\dot \Porb$ at $t = T_0$ shown in Table~\ref{tab1},
with a $\chi^2 = 2.2$ (for 1 d.o.f.). This corresponds to a probability of
$13\%$ of obtaining a $\chi^2$ larger than the one we found. Our result 
is therefore acceptable (or, better, not rejectable), since the probability 
we obtained is above the conventionally accepted significance level of  $5\%$ 
(in this case, in fact, the discrepancy between the expected and observed values
of $\chi^2$ is not significant since the two values are within less than 
$2\; \sigma$ from each other, see e.g.\ Bevington \& Robinson 2003).  We find a
highly significant derivative of the orbital period, which
indicates that the orbital period in this system is increasing at a rate
of $(3.40 \pm 0.12) \times 10^{-12}$ s/s.
Note that H08, simultaneously and independently, found a very similar result 
for the orbital period derivative in \saxj, $\dot P_{orb} = (3.5 \pm 0.2) 
\times 10^{-12}$ s/s, compatible with the result reported in this paper. 
The only difference is in the quoted error. H08 state that this difference 
is possibly due to an underestimate of the phase uncertainties reported in 
this paper, as testified by a worse reduced $\chi^2$. 
We just note that we do not have any evidence that the phase uncertainties 
we derive are underestimated. Indeed our reduced $\chi^2$ is worse than 
the one obtained by H08, that is $\chi^2 = 1.01$ for 1 dof), but it is still 
statistically acceptable. 
However, in the hypothesis that our errors are slightly underestimated, 
we have increased by a factor of 1.5 all the errors on the phases in order to
obtain a $\chi^2$ as close as possible to 1. 
In this way we obtain a $\chi^2$ of 0.98 for 1 d.o.f., and we have 
re-evaluated the errors on the orbital parameters, finding that these errors
increase by a factor of 1.5. This means that our "conservative" estimate of
the orbital period derivative is $(3.40 \pm 0.18) \times 10^{-12}$ s/s 
($90\%$ c.l.).
As a final check we have fitted with the same formula the points shown in the 
inset of Figure~\ref{fig1}, obtaining, as expected, the same results.

Note that the orbital period of \saxj\ is now known with a precision of
1 over $10^9$, that is an improvement of one order of magnitude respect to
the previous estimate by Papitto et al.\ (2005) and two orders of 
magnitude with respect to Chakrabarty \& Morgan (1998). On the other hand, 
the derivative of the orbital period indicates that the orbital period in 
this system is increasing at the quite large rate of $(3.40 \pm 0.18) 
\times 10^{-12}$ s/s, that is at least an order of magnitude higher than 
what is predicted by a conservative mass transfer driven by Gravitational
Radiation (GR, see below). 
In the next section we discuss a possible explanation for this surprising result.
\setcounter{table}{0}
\begin{table*}
  \begin{minipage}{80mm}
    \caption{Best fit orbital parameters for \saxj. }
    \label{tab1}
    \begin{tabular}{@{}lr}
      \hline
Parameter & Value  \\
      \hline
$T^*_0$ (days)   	& 50914.8784320(11) \\
$P_{orb}$ (s)    	& 7249.156499(18) \\
$\dot P_{orb}$ (s/s)	& $3.40(18) \times 10^{-12}$ \\
$a_1 \sin i $ (lt-ms)   & $62.809(1)^a$ \\
$e$                     & $< 4.6 \times 10^{-5}$ \\
$\chi^2 / dof$		& $0.98 / 1$ \\
\hline
    \end{tabular}
    \medskip

The reference time at which the orbital period, $\Porb$ and its derivative,
$\dot \Porb$, are referred to is the beginning of the 1998 outburst, that is 
$T_0 = 50914.8099$ MJD. Numbers in parentheses are the uncertainties 
in the last significant digits at $90\%$ c.l. Upper limits are at $95\%$ c.l. 
Uncertainties are calculated conservatively increasing the errors on the 
fitted points in order to reach a $\chi^2 / dof \simeq 1$, as described 
in the text.\\
$^a$ The value of $a_1 \sin i$ and its 1-$\sigma$ error are from 
Chakrabarty \& Morgan 1998.
  \end{minipage}
\end{table*}

\section{Discussion}

We have performed a precise timing analysis of all the X-ray outbursts 
of the AMSP \saxj\ observed to date by the RXTE/PCA, covering more
than 7 years in time. We divided each outburst in several intervals and found, 
for each interval, differential corrections to previously published orbital 
parameters. 
The obtained times of passage of the NS at the ascending node were significantly 
different in different outbursts. We fitted these times with a parabolic function 
of time finding an improved orbital solution valid over a time span of more than 7 
years. This solution includes a highly significant derivative of the 
orbital period, $\dot P_{orb} = (3.40 \pm 0.18) \times 10^{-12}$ s/s.
This value, found simultaneously and independently by H08, is the first measure 
of the orbital period derivative for an AMSP. 
However, this orbital period derivative is quite unexpected, since it is more 
than one order of magnitude higher than what is expected from conservative mass 
transfer driven by GR.

Orbital evolution calculations show that the orbital period derivative caused by 
conservative mass transfer induced by emission of GR is given by: 
\begin{equation}
\dot P_{\rm orb} = -1.4 \times 10^{-13} \; m_1 \; m_{2, \; 0.1} \; m^{-1/3} 
P_{2h}^{-5/3} \; [(n - 1/3) / (n + 5/3 - 2q)]\;s s^{-1} 
\label{eq1}
\end{equation} 
(derived from Verbunt 1993; see also Rappaport et al.\ 1987), 
where $m_1$ and $m$ are, respectively, the mass of the primary, $M_1$, and 
the total mass, $M_1 + M_2$, in units of solar masses, $m_{2, \; 0.1}$ is 
the mass of the secondary in units of $0.1\; M_\odot$, $P_{2h}$ is the orbital 
period in units of 2~h, $q = m_2/m_1$ is the mass ratio and where $n$ is the 
index of the mass-radius relation of the secondary $R_2 \propto M_2^{\rm n}$. 
Therefore a positive orbital period derivative certainly indicates a 
mass-radius index $n < 1/3$, and therefore, most probably, a degenerate 
or fully convective companion star (see e.g.\ King 1988). However,
the $\dot P_{\rm orb}$ we measure is an order of magnitude higher than 
what is expected from GR. 

\subsection{Conservation of angular momentum}
In order to explain the quite unexpected value measured for the orbital period 
derivative, we start from the equation of the angular momentum of the system, 
which must be verified instantaneously. The orbital angular momentum of the system
can be written as: $J_{\rm orb} = [G a / (M_1 + M_2)]^{1/2} M_1 M_2$, 
where $G$ is the Gravitational constant, and $a$ is the orbital separation.
We can differentiate this expression in order to find the variation of the orbital
angular momentum of the system caused by mass transfer. We indicate with 
$- \dot M_2$ the mass transferred by the secondary, which can be accreted 
onto the neutron star (conservative mass transfer) or can be lost from the system
(non-conservative mass transfer). We can therefore write $\dot M_1 = - \beta 
\dot M_2$, where $\beta$ is the fraction of the transferred mass that is 
accreting onto the neutron star, while $1-\beta$ is the fraction of the 
transferred mass that is lost from the system. The specific angular momentum,
$l_{\rm ej}$, with which the transferred mass is lost from the system can be 
written in units of the specific angular momentum of the secondary, that is:
$\alpha = l_{ej} / (\Omega_{\rm orb} r_2^2) = 
l_{ej} P_{orb} (M_1 + M_2)^2/(2 \pi a^2 M_1^2)$, where $r_2$ is the distance 
of the secondary star from the center of mass of the system. Calculating the
derivative of the orbital angular momentum of the system, we obtain:
\begin{equation}
\label{dotP}
\frac{\dot P_{\rm orb}}{P_{\rm orb}} = 3 \left[\frac{\dot J}{J_{\rm orb}} - 
\frac{\dot M_2}{M_2} \; g(\beta,q,\alpha)\right],
\end{equation}
where $g(\beta,q,\alpha) = 1 - \beta q - (1-\beta) (\alpha + q/3)/(1+q)$, and
$\dot J / J_{\rm orb}$ represents possible losses of angular momentum from the system
(e.g.\ caused by GR). Since the term $\dot J / J_{\rm orb}$
must be negative, while the measured $\dot P_{\rm orb} / P_{\rm orb}$ is positive,
we can derive a lower limit on the positive quantity $-\dot M_2 / M_2$ assuming 
that $\dot J / J_{\rm orb} = 0$:
\begin{equation}
\label{dotPlimit}
\frac{\dot P_{\rm orb}}{P_{\rm orb}} \le 3 \left(- \frac{\dot M_2}{M_2} \;
g(\beta,q,\alpha)\right).
\end{equation}

Assuming a conservative mass transfer, $\beta = 1$, it is easy to see that 
$g(1,q,\alpha) = 1-q \simeq 1$, where we have used the information that for
\saxj\ the mass function gives $q \ge 4 \times 10^{-2}$ for $M_1 = 1.4\; 
M_\odot$ (Chakrabarty \& Morgan 1998), and can be therefore neglected. 
Hence, for conservative mass transfer the conservation of angular 
momentum gives: 
$\dot P_{\rm orb}/P_{\rm orb} \le 3 (-\dot M_2 / M_2)$. We can estimate the 
averaged mass transfer rate from the averaged observed luminosity of the 
source. Since \saxj\ accretes for about 30 days every two years, we have 
estimated an order of magnitude for the averaged X-ray luminosity from the 
source, that is $L_X \sim 4 \times 10^{34}$ ergs/s, which corresponds to
$3 (-\dot M_2 / M_2) = 6.6 \times 10^{-18}$ s$^{-1}$. It is easy to see that
the measured $\dot P_{\rm orb}/P_{\rm orb}$, $4.7 \times 10^{-16}$ s$^{-1}$, 
is at least 70 times higher than the value predicted in the conservative 
mass transfer case, hence excluding this scenario.

Assuming a totally non-conservative mass transfer, $\beta = 0$, we find that
$g(0,q,\alpha) = (1 - \alpha + 2 q / 3)/(1+q) \simeq 1-\alpha$, implying that
$\dot P_{\rm orb}/P_{\rm orb} \le 3 (1-\alpha) (-\dot M_2 / M_2)$. Since
the first term is positive we find that $\alpha < 1$ and the specific angular 
momentum with which matter is expelled from the system must be less than
the specific angular momentum of the secondary. For matter leaving the 
system with the specific angular momentum of the primary we have 
$\alpha = q^2 \sim 0$. In this case we find, therefore, the same result 
of the conservative case where no angular momentum losses from the system 
occur. 
This is due to the fact that the specific angular momentum of the primary
is so small that there is no difference with respect to the conservative
case. Since the specific angular momentum of the mass lost must be in between
the specific angular momentum of the primary and that of the secondary,
a reasonable hypothesis is that matter leaves the system with the specific
angular momentum of the inner Lagrangian point. In this case
$\alpha = [1-0.462 (1+q)^{2/3} q^{1/3}]^2 \simeq 0.7$,
where we have used for the Roche Lobe radius the approximation given by
Paczy\'nski (1971). We therefore find $\dot P_{\rm orb}/P_{\rm orb} \le 
(-\dot M_2 / M_2)$. Using the measured value of the quantity 
$\dot P_{\rm orb}/P_{\rm orb} = 4.7 \times 10^{-16}$ s$^{-1}$, we find 
that to explain this result in a totally non-conservative scenario the 
mass transfer rate must be: 
$(-\dot M_2) = \dot M_{\rm ej} \ge 8.3 \times 10^{-10}\; M_\odot\; yr^{-1}$.
We can therefore explain the measured derivative of the orbital period
of the system assuming that the system is expelling matter at a quite
large rate, that may be as high as $\sim 10^{-9}\; M_\odot\; yr^{-1}$, and this
is found just assuming the conservation of the angular momentum of the
system, and independently of the secular evolution adopted. 

\subsection{Possible secular evolution of the system}
In order to verify whether this result is just a transient peculiar behavior 
of the system due to unknown causes or it is instead compatible 
with a secular evolution, we have solved the secular evolution equations
of the system using the assumptions described in the following.
i) Angular momentum losses are due to GR and are given by:
$\dot J / J = -32 G^3 M_1 M_2 (M_1 + M_2) / (5 c^5 a^4)$, where 
$c$ is the speed of light, 
and $J = M_1 M_2 [G a / (M_1 + M_2)]^{1/2}$ is the binary angular momentum 
(see e.g.\ Landau \& Lifschitz 1958; Verbunt 1993). ii) For the secondary 
we have adopted a mass-radius relation $R_2 \propto M_2^n$. 
iii) We have imposed that the radius of the secondary follows the evolution of 
the secondary Roche Lobe radius: 
$\dot R_{L2} / R_{L2} = \dot R_2 / R_2$, where 
for the radius of the secondary Roche Lobe we have adopted the Paczy\'nski (1971)
approximation: $R_{L2} = 2/3^{4/3} [q/(1+q)]^{1/3} a$ that is valid for small
mass ratios, $q = M_2 / M_1 \le 0.8$.
In these hypotheses we derived a simple expression for the orbital period 
derivative and the mass transfer rate in the extreme cases of totally conservative
and totally non-conservative mass transfer (see e.g.\ Verbunt 1993; van Teeseling 
\& King 1998; King et al.\ 2003; King et al.\ 2005):
\begin{equation}
\dot P_{orb} = -1.38 \times 10^{-12} 
\left[\frac{n - 1/3}{n - 1/3 + 2 g} \right]
m_1^{5/3} q (1+q)^{-1/3} P_{2h}^{-5/3} \;\; {\rm s/s} 
\label{eqdotporb}
\end{equation}
\begin{equation}
\dot M = -\dot M_2 = 4.03 \times 10^{-9} 
\left[\frac{1}{n - 1/3 + 2 g} \right] 
m_1^{8/3} q^2 (1+q)^{-1/3} P_{2h}^{-8/3} \;\; {\rm M_\odot / yr} 
\label{eqdotm}
\end{equation}
where $g = 1-q$ for totally conservative mass transfer (in this case it is 
easy to see that eq.~\ref{eqdotporb} gives the same $\dot P_{orb}$ given by 
eq.~\ref{eq1}, as expected), and $g = 1 - (\alpha + q/3) / (1+q)$ for totally 
non-conservative mass transfer.

Comparing the measured orbital period derivative with eq.~\ref{eqdotporb}
(assuming that the orbital period derivative we measure reflects the secular 
evolution of the system), we note that, in order to have $\dot P_{orb} > 0$ 
we have to assume an index $n < 1/3$.
In the case of \saxj\ the secondary mass is $m_2 \le 0.14$ at 95\% 
c.l. (Chakrabarty \& Morgan 1998), and therefore the mass ratio $q \le 0.1$ 
implies that for the totally conservative mass transfer case ($g=1-q$), the 
$\dot P_{orb}$ expected from GR must be of the order of $10^{-13}$ s/s, not 
compatible with what we measured for \saxj. 
In other words, if we assume a conservative mass transfer for the system (that 
means that the mass transferred during outbursts is completely accreted by the NS, 
and during quiescence no or negligible mass is accreted or lost from the system), 
we find that it is impossible to explain the observed orbital period 
derivative with a secular evolution driven by GR.

\begin{figure}
  \begin{center}
    \begin{minipage}{80mm}
      \includegraphics[width=70mm]{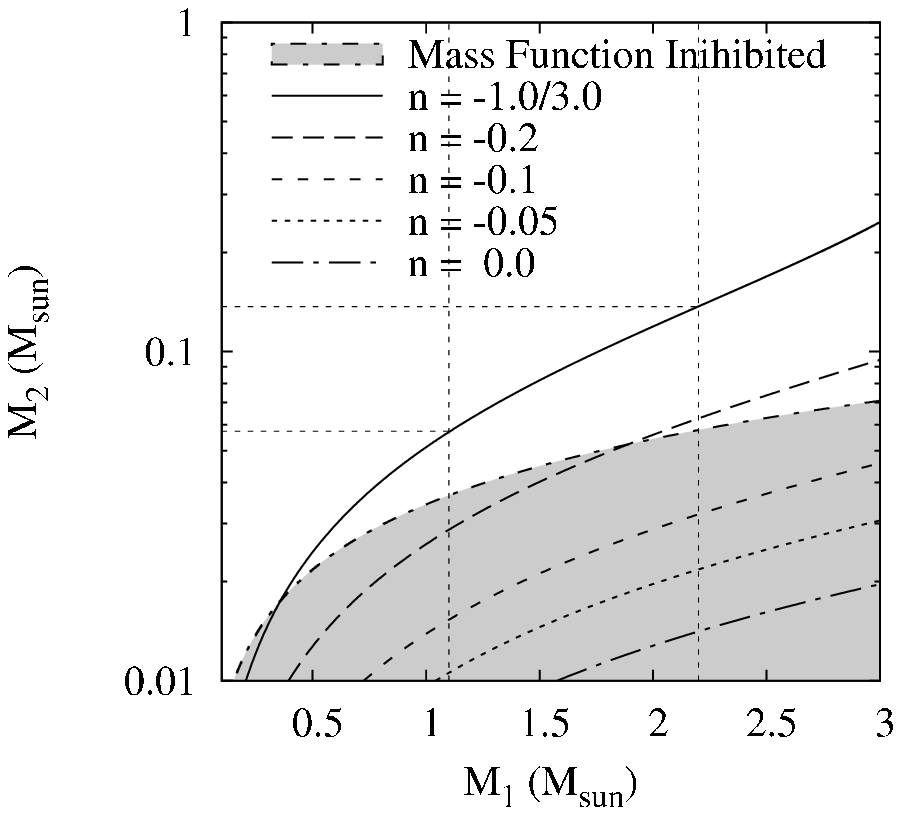}
      \includegraphics[width=70mm]{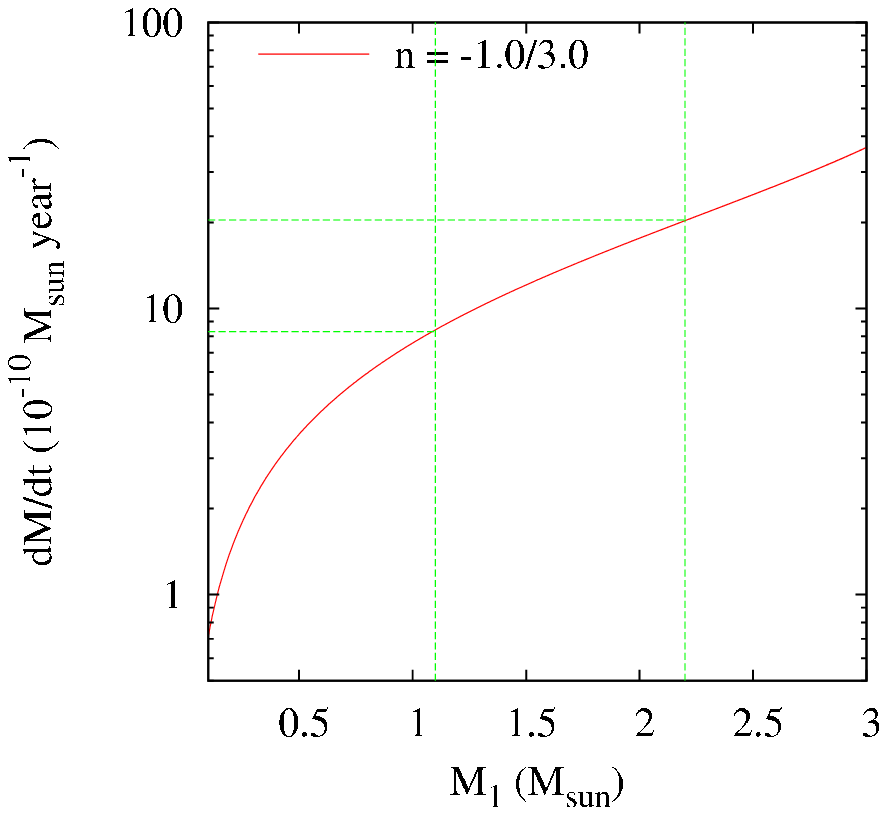}
      \caption{{\bf Top:} Companion star mass vs.\ NS mass in the hypothesis of 
       totally non conservative mass transfer (with matter leaving the system with 
       the specific angular momentum at the inner Lagrangian point) and assuming 
       the $\dot P_{orb}$ measured for \saxj. Different curves correspond to 
       different values of the mass-radius index $n$ of the secondary. 
       Horizontal lines indicate the limits for the secondary star mass 
       corresponding to reasonable limits for
       the NS mass and to $n=-1/3$.
       {\bf Bottom:} Mass rate outflowing the secondary Roche Lobe in the 
       hypothesis of totally non conservative mass transfer (as above) and 
       assuming $n=-1/3$.}
    \label{fig2} 
    \end{minipage}
  \end{center}
\end{figure}

On the other hand, if we assume that during X-ray quiescence the companion star
is still overflowing its Roche Lobe but the transferred mass is not 
accreted onto the NS and is instead ejected from the system, we find a good 
agreement between the measured and expected orbital period derivative assuming 
that the matter leaves the system with the specific angular momentum at the inner 
Lagrangian point, $\alpha = [1 - (2/3^{4/3}) q^{1/3} (1+q)^{2/3}]^2$.
Adopting the measured value $\dot P_{orb} = 3.4 \times 10^{-12}$ s/s and 
the other parameters of \saxj, eq.~\ref{eqdotporb} translates into a relation 
between $m_1$, $m_2$ and the mass-radius index $n$; this is plotted in 
Figure~\ref{fig2} (top panel) for different values of $n$ going from 0 to 
$n = -1/3$. 
The constraint on $m_1$ vs.\ $m_2$ imposed by the mass function of the system 
is also plotted (the shadowed region in the figure) and indicates that the most 
probable value for $n$ is $-1/3$, which in turn indicates a degenerate or, 
most probably, a fully convective companion star.  
In fact, in a system with orbital period less than 3 h, where the
mass of the Roche-Lobe filling companion is below $0.2-0.3\; \msun$, 
the companion star becomes fully convective with a mass-radius hydrostatic 
equilibrium equation $R \propto M^{-1/3}$ (e.g.\ King 1988; Verbunt 1993).
Also, for reasonable minimum, average and maximum values of the NS mass, 
1.1,  1.56, and  2.2 $\msun$, respectively, we obtain the following values 
for the secondary mass: 
0.053, 0.088, and 0.137~$M_\odot$, and the following values for the 
inclination of the system: 44$^\circ$, 32$^\circ$, and 26$^\circ$, respectively.

Assuming therefore $n = -1/3$ we have plotted in Figure~\ref{fig2} (bottom panel) 
the corresponding non-conservative mass transfer rate as a function of $m_1$. 
We find that for $m_1 = 1.5$ the mass transfer rate must be of the order of 
$10^{-9}\; M_\odot$/yr, much higher than what is expected in a conservative 
GR driven mass transfer case. 
Note that this high $\dot M$ might explain the spin period evolution reported 
by Burderi et al. (2006; see, however, H08 who could not detect a spin period 
derivative during the outbursts).
Actually, during the X-ray outbursts, the mass transfer is conservative since
the transferred matter is accreted onto the NS. However, the accretion phase
duty cycle, about 40 days / 2 years = $5\%$, is so small that the totally
non-conservative scenario proposed above is a good approximation.

\subsection{Is \saxj\ a 'hidden' black widow pulsar?}
If the hypothesis of a highly non-conservative mass transfer in \saxj\ is
correct, the question to answer is why is accretion inhibited during 
X-ray quiescence while the companion star is transferring mass at a high rate? 
We propose that the answer has to be found in the radiation pressure of the 
magneto-dipole rotator emission, with a mechanism that is similar to what 
is proposed to explain the behavior of the so-called black widow pulsars 
(see e.g.\ Tavani et al. 1991a; King et al. 2003; 2005; see also Burderi et 
al.\ 2001). Indeed, the possibility that the magneto-dipole emission is 
active in \saxj\ during X-ray quiescence has been invoked by 
Burderi et al.\ (2003, see similar results in Campana et al.\ 2004) to 
explain the optical counterpart of the source, which is observed to be 
over-luminous during quiescence (Homer et al.\ 2001). 
In this scenario, the optical luminosity in quiescence is explained by the 
spin-down luminosity of the magneto-dipole rotator (with a magnetic 
field of $(1-5) \times 10^8$ Gauss) which is reprocessed by 
the companion star and/or a remnant accretion disc. Interestingly,
similar evidence of a strongly irradiated companion star during quiescence 
has been found also for \igrj, the fastest among the known AMSPs 
(D'Avanzo et al.\ 2007).

In other words, a temporary reduction of the mass-accretion rate onto
the neutron star (note that the so-called disc Instability Model, DIM --
see e.g.\ Dubus, Hameury, \& Lasota 2001 -- usually invoked to explain 
the transient behavior of these sources, may play a role in triggering 
or quenching the X-ray outbursts in \saxj) may cause the switch on of 
the emission of the magneto-dipole rotator, and, in some cases, even 
if the mass transfer rate has not changed, the accretion of 
matter onto the NS can be inhibited because the radiation pressure from 
the radio pulsar may be capable of ejecting out of the system most of the 
matter overflowing from the companion (see e.g.\ Burderi et al.\ 2001 
and references therein).  This phase has been termed ``radio--ejection''.  
One of the strongest predictions of this model is the presence, 
during the radio-ejection phase, of a strong wind of matter emanating 
from the system: the mass overflowing from the companion swept away 
by the radiation pressure of the pulsar. Indeed, the existence of 
'hidden' millisecond pulsars, whose radio emission is completely 
blocked by material engulfing the system that is continuously replenished 
by the mass outflow driven by companion irradiation, has already been 
predicted by Tavani (1991a).

A beautiful confirmation of this model was provided by the discovery
of PSR J1740--5340, an eclipsing millisecond radio pulsar, with a spin 
period of 3.65 ms, located in the globular cluster NGC 6397 (D'Amico 
et al. 2001).  It has the longest orbital period ($P_{\rm orb} \simeq 
32.5$ hrs) among the 10 eclipsing pulsars detected up to now.  
The peculiarity of this source is that the companion is a slightly 
evolved turnoff star still overflowing its Roche lobe. This is 
demonstrated by the presence of matter around the system that causes 
long lasting and sometimes irregular radio eclipses, and by the shape 
of the optical light curve, which is well modeled assuming a Roche-lobe 
deformation of the mass losing component (Ferraro et al. 2001).  
An evolutionary scenario for this system has been proposed by Burderi, 
D'Antona, \& Burgay (2002), who provided convincing evidence that
PSR J1740--5340 is an example of a system in the radio-ejection phase, 
by modeling the evolution of the possible binary system progenitor.
In other words, PSR J1740--5340 can be considered a 'star-vaporizing 
pulsar' of type II in the terminology used by Tavani (1991a).

We believe that the behavior of \saxj\ is very similar to the one of
PSR J1740--5340, the main differences being the orbital period, which is 
$\sim 32$ h in the case of PSR J1740--5340 and $\sim 2$ h in the case of 
\saxj, and the mass transfer rate from the companion, which has been
estimated to be $\sim 10^{-10} \; M_\odot$ yr$^{-1}$ for PSR J1740--5340
and one order of magnitude higher for \saxj. Both these factors will 
increase the local Dispersion Measure (DM) to the source in the case of 
\saxj, and hence will predict a much higher free-free absorption in 
the case of \saxj. This is in agreement with the fact that, although 
widely searched, no radio pulsations have been detected from \saxj\ up 
to date (Burgay et al.\ 2003). In other words, \saxj\ can be considered 
a 'hidden' millisecond pulsar, or a 'star-vaporizing pulsar' of 
type III in the terminology used by Tavani (1991a).

A similar highly non conservative mass transfer, triggered by irradiation
of the secondary and/or of an accretion disc by the primary (according to the
model of Tavani et al.\ 1991b), has been proposed to explain the large
orbital period derivative observed in the ultra-compact Low Mass X-ray 
Binary X~1916--053, composed of a neutron star and a semi-degenerated 
white dwarf, exhibiting periodic X-ray dips. In this case, 
$\dot{P}_{orb}/P_{orb} \simeq 5.1 \times 10^{-15}\; s^{-1}$, which implies
a mass transfer rate of $\sim 10^{-8}\; M_\odot$ yr$^{-1}$, with $60-90 \%$
of the companion mass loss outflowing from the system (Hu, Chou, \& Chung
2008).



As predicted by several authors (e.g.\ Chakrabarty \& Morgan 1998; 
King et al.\ 2005), and in agreement with our interpretation of the 
orbital period derivative in \saxj\ as due to a highly non-conservative 
mass transfer, we propose therefore that \saxj, and other similar 
systems, belong to the population of the so-called black widow pulsars 
(or are evolving towards black widow pulsars); these are millisecond 
radio pulsars thought to ablate the companion and likely able to 
produce large mass outflows. 
When (or if) the pressure of the outflowing matter becomes
sufficiently high to temporarily overcome the radiation pressure of
the magneto-dipole rotator, the source experiences a transient mass
accretion episode, resulting in an X-ray outburst.
Indeed, \saxj\ and the other known AMSPs are all transient systems 
(accreting just for a very short fraction of the time), with small values 
of the mass function (implying small minimum mass for the secondary) and 
short orbital periods (less than a few hours).\footnote{Among the presently
known AMSPs, the newly discovered intermittent pulsar
SAX J1748.9--2021 (Altamirano et al.\ 2008) in the globular cluster NGC 6440
is an exception since this pulsar shows mass function and orbital period 
higher than the other Galactic AMSP. However, it is worth noting that 
this pulsar belongs to a globular cluster in which capture in the dense 
cluster environment may have played a role.} 

Although in some of these black widow radio pulsars (variable) radio eclipses 
have been observed, 
clearly demonstrating the presence of matter around the system, a direct proof 
of severe mass losses from these system has never been found to date.
The orbital evolution of \saxj\ indicates that this X-ray transient 
millisecond pulsar indeed may expel mass from the system for most of the time 
with just short episodes of accretion observed as X-ray outbursts. We therefore  
propose that \saxj\ (and perhaps most AMSPs) is indeed a black widow still eating
the companion star.

However, some black widow pulsars have shown quite complex derivatives of the
orbital period. In particular, the prototype of this class, PSR B1957+20 
(a type-I star-vaporizing eclipsing millisecond pulsar), shows a large first 
derivative of the orbital period (almost an order of magnitude higher than 
the one of \saxj), and also a second orbital period derivative, indicating 
a quasi-cyclic orbital period variation (Arzoumanian, Fruchter, \& Taylor
1994). A similar behavior has been observed in another binary millisecond 
pulsar, PSR J2051--0827, which shows a third derivative of the orbital 
period and a variation of $a \sin i$, indicating that the companion is 
underfilling its Roche lobe by $\sim 50\%$ (Doroshenko et al.\ 2001). 
In these cases, the complex orbital period variation has been ascribed to
gravitational quadrupole coupling (i.e.\ a variable quadrupole moment of
the companion which is due to a cyclic spin-up and spin-down of the star 
rotation). In this scenario, the companion star must be partially 
non-degenerate, convective and magnetically active, so that the wind of 
the companion star will result in a strong torque which tends to slow down 
the star, making the companion star rotation deviating from synchronous 
rotation ($\Omega_C = f 2 \pi /P_{\rm orb}$, with $f < 1$).

In the case of \saxj, the present data do not allow to find a 
second derivative of the orbital period, and therefore it is not clear if
the orbital period derivative will change sign. However, we note that
for a (type-I) black widow (radio) pulsar the companion star may be no 
more strongly orbitally locked; hence deviations from 
co-rotation may be possible, and this may reflect in strange (cyclic) 
changes of the orbital period of the system. However, in the case of \saxj, 
that is an accreting neutron star, the companion must be completely 
orbitally locked (since the secondary star still fills its Roche lobe
during X-ray outbursts and cannot detach in a time scale of a few years) 
and therefore the most probable way to change the orbital period of the 
system is a change of the averaged specific angular momentum, which can be  
obtained transferring mass from the secondary to the neutron star or 
expelling mass from the system with appropriate specific angular momentum, 
less than the specific angular momentum of the secondary.

In conclusion, we propose that we are witnessing the behavior of a 'hidden' 
black widow, eating its companion during X-ray outbursts and ablating it 
during quiescence; the next X-ray outburst of \saxj\ will be of 
fundamental importance to test or disprove this scenario. 
Moreover, the analogy of \saxj\ with a black widow should also be 
tested observationally; observing the source at other wavelength may give
important information. For instance, it is already known that \saxj\ shows 
transient radio emissions; this was observed for the first time at the end of 
the 1998 outburst, approximately 1 day after the onset of a rapid decline in 
the X-ray flux, by Gaensler, Stappers, \& Getts (1999) and was attributed to 
an ejection of material from the system. The possible presence of an $H \alpha$ 
bow shock nebula (like the one observed in PSR B1957+20) may be difficult to 
test in this case, given the crowding around the source in the optical band 
(see Campana et al.\ 2004), while it may be important to look for an IR excess 
which may be caused by excess of matter around the system.

\thanks{We thank A. King for useful discussions, and the second referee of this
paper for useful comments which helped us to improve the manuscript. 
We acknowledge the use of RXTE data from the HEASARC public archive. This work 
was supported by the Ministero della Istruzione, della Universit\`a e 
della Ricerca (MIUR), national program PRIN2005 2005024090$\_$004.}


\end{document}